\documentclass{article}

\usepackage{amsmath,amsfonts,amssymb,times,graphicx,natbib,algorithm,algorithmic,hyperref}

\DeclareMathOperator*{\argmin}{arg\:min}
\usepackage[accepted]{data4good2016}

\usepackage{subfigure} 

\usepackage[capitalize]{cleveref}

\usepackage{booktabs}
\usepackage[multiple]{footmisc}

\icmltitlerunning{Cyberbullying Identification Using Participant-Vocabulary Consistency}

\begin{document}

\twocolumn[
\icmltitle{Cyberbullying Identification Using Participant-Vocabulary Consistency}

\icmlauthor{Elaheh Raisi}{elaheh@vt.edu}
\icmladdress{Virginia Tech, Blacksburg, VA}
\icmlauthor{Bert Huang}{bhuang@vt.edu}
\icmladdress{Virginia Tech, Blacksburg, VA}

\vskip 0.3in
]

\begin{abstract}
With the rise of social media, people can now form relationships and communities easily regardless of location, race, ethnicity, or gender. However, the power of social media simultaneously enables harmful online behavior such as harassment and bullying. Cyberbullying is a serious social problem, making it an important topic in social network analysis. Machine learning methods can potentially help provide better understanding of this phenomenon, but they must address several key challenges: the rapidly changing vocabulary involved in cyberbullying, the role of social network structure, and the scale of the data. In this study, we propose a model that simultaneously discovers instigators and victims of bullying as well as new bullying vocabulary by starting with a corpus of social interactions and a seed dictionary of bullying indicators. We formulate an objective function based on participant-vocabulary consistency. We evaluate this approach on Twitter and Ask.fm data sets and show that the proposed method can detect new bullying vocabulary as well as victims and bullies.
\end{abstract}

\section{Introduction}
\label{sec:intro}

\begin{figure}[tb]
\begin{center}
\centerline{\includegraphics[height = 2 in,width=\columnwidth]{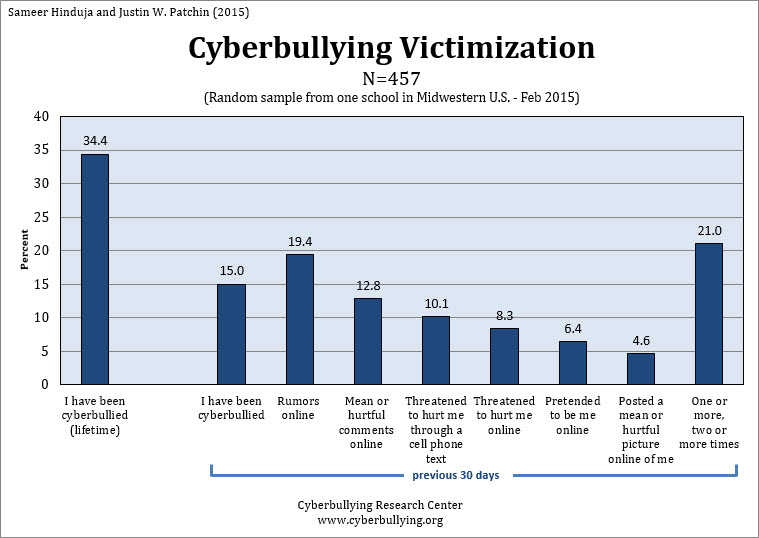}}
\vspace{-1em}
\caption{Survey Statistics on Cyberbullying Experiences. Data collected and visualized by the Cyberbullying Research Center (http://cyberbullying.org/).}
\vspace{-2em}
\label{icml-historical}
\end{center}
\vspace{-1.5em}
\end{figure}

Social media has significantly changed the nature of society. Our ability to connect with others has been massively enhanced, removing boundaries created by location, gender, age, and race. However, the benefits of this hyper-connectivity also come with the enhancement of detrimental aspects of social behavior. Cyberbullying is an example of one such behavior that is heavily affecting the younger generations \citep{boyd:14}. The Cyberbullying Research Center defines cyberbullying as ``willful and repeated harm inflicted through the use of computers, cell phones, and other electronic devices.'' Like traditional bullying, cyberbullying occurs in various forms. Examples include name calling, rumor spreading, threats, and sharing of private information or photographs.\footnote{http://www.endcyberbullying.org}\textsuperscript{,}\footnote{http://www.ncpc.org/cyberbullying} Even seemingly innocuous actions such as supporting offensive comments by ``liking'' them can be considered bullying \citep{wang:2009}. As stated by the National Crime Prevention Council, around 50\% of American young people are victimized by cyberbullying.
According to the American Academy of Child and Adolescent Psychiatry, victims of cyberbullying have strong tendencies toward mental and psychiatric disorders  \citep{aaca}. In extreme cases, suicides have been linked to cyberbullying \citep{Phoebe, hannah}. The phenomenon is widespread, as indicated in \cref{icml-historical}, which plots survey responses collected from students. These facts make it clear that cyberbullying is a serious health threat.

Machine learning can be useful in addressing the cyberbullying problem.
Recently, various studies considered supervised, text-based cyberbullying detection, classifying social media posts as \lq bullying \rq or \lq non-bullying\rq. Training data is annotated by experts or crowdsourced workers. Since bullying often involves offensive language, text-based cyberbullying detection studies often use curated swear words as features, augmenting other standard text features. Then, supervised machine learning approaches train classifiers from this annotated data \citep{yin:2009, Ptaszynski:2010, Dinakar:2011}. 

We identify three significant challenges for supervised cyberbullying detection. First, annotation is not an easy task. It requires expertise about culture, examination of the social structure of the individuals involved in each interaction. Because of the difficulty of labeling these often subtle distinctions between bullying and non-bullying, there is likely to be disagreement among labelers, making costs add up quickly for a large-scale problem. Second, reasoning about which individuals are involved in bullying should do joint, or collective, classification. E.g., if we believe a message from A to B is a bullying interaction, we should also expect a message from A to C to have an increased likelihood of also being bullying. Third, language is rapidly changing, especially among young populations, making the use of static text indicators prone to becoming outdated. Some curse words have completely faded away or are not as taboo as they once were, while new slang is frequently introduced into the culture. These three challenges suggest that we need a dynamic methodology to collectively detect emerging and evolving slurs with only weak supervision.

In this paper, we introduce an automated, data-driven method for cyberbullying identification. The eventual goal of such work is to detect such harmful behaviors in social media and intervene, either by filtering or by providing advice to those involved. Our proposed learnable model takes advantage of the fact that the data and concepts involve relationships. We train this relational model in a weakly supervised manner, where human experts provide a small seed set of phrases that are highly indicative of bullying. Then the algorithm finds other bullying terms by extrapolating from these expert annotations. In other words, our algorithm detects cyberbullying from key-phrase indicators. We refer to our proposed method as the participant-vocabulary consistency (PVC) model; It seeks a consistent parameter setting for all users and key phrases in the data that characterizes the tendency of each user to harass or to be harassed and the tendency of a key phrase to be indicative of harassment. The learning algorithm optimizes the parameters to minimize their disagreement with the training data which are highly indicative bullying phrases in messages between specific users.

A study by \citet{survey:13} found that Facebook, YouTube, Twitter, and Ask.fm are the platforms that have the most frequent occurrences of cyberbullying. To evaluate the participant-vocabulary consistency method, we ran our experiments on Twitter and Ask.fm data. From a list of highly indicative of bullying key phrases, we subsample small seed sets to train the algorithm. We then examine the participant-vocabulary consistency method to see how well it recovers the remaining, held-out set of indicative phrases. Additionally, we extract the detected most bullying phrases and qualitatively verify that they are in fact examples of bullying.

\section{Related Work}
There are two main branches of research related to our topic. One of them is online harassment and cyberbullying detection; the other one is associated with automated vocabulary discovery. 
Various studies have used fully supervised learning to classify bully posts from non-bully posts. Many of them focus on the textual features of post to identify cyberbullying incidents \citep{Dinakar:2011, Ptaszynski:2010, hosseinmardi:aaai15, Chen-SocialCom:2012, Margono:2014}.
Some of them use other features than only textual features, for example content, sentiment, and contextual features \citep{ yin:2009}, the number, density and the value of offensive words \citep{ Reynolds:2011}, or the number of friends, network structure, and relationship centrality \citep{Huang:2014}.  \citet{nahar:2013} used semantic and weighted features; they also identify predators and victims using a ranking algorithm. 
Many studies have been applied machine learning techniques to better understand social-psychological issues such as bullying. They used data sets such as Twitter, Instagram and Ask.fm to study negative user behavior \citep{Bellmore:2015, hosseinmardi:asonam14, hosseinmardi:BdCloud14}.

Various works use query expansion to extend search queries to dynamically include additional terms. For example, \citet{Massoudi:2011} use temporal information as well as co-occurrence to score the related terms to expand the query.
\citet{Mahendiran_discoveringevolving} propose a method based on probabilistic soft logic to grow a vocabulary using multiple indicators (e.g., social network, demographics, and time).

\section{Proposed Method}

To model the cyberbullying problem, for each user $u_i$, we assign a bully score $b_i$ and a victim score $v_i$. The bully score measures how much a user tends to bully others; likewise, victim score indicates how much a user tends to be bullied by other users. For each feature $w_k$, we associate a feature-indicator score that represents how much the feature is an indicator of a bullying interaction. Each feature represents the existence of some descriptor in the message, such as n-grams in text data. 
The sum of sender’s bullying score and receiver’s victim score ($b_i + v_i$) specifies the message's \emph{social bullying score}, which our model aims to make consistent with the vocabulary-based feature score. We formulate a regularized objective function that penalizes inconsistency between the social bullying score and each of the feature scores.
\begin{equation}
\begin{aligned}
&J(\mathbf{b}, \mathbf{v}, \mathbf{w}; \lambda) = ~ \frac{\lambda}{2} \left( ||\mathbf{b}||^2 + ||\mathbf{v}||^2 + ||\mathbf{w}||^2 \right) + \\
& ~ \frac{1}{2} \sum_{m \in M} \left( \sum_{k : w_k \in f(m)} \left( b_{s(m)} + v_{r(m)} - w_k \right)^2 \right)
\label{eq:objective}
\end{aligned}
\end{equation}
Learning is then an optimization problem over parameter vectors $\mathbf{b}$, $\mathbf{v}$, and $\mathbf{w}$. The consistency penalties are determined by the structure of the social data. We include information from an expert-provided initial seed of highly indicative bully words. We require these seed features to have a high score, adding the constraint:
\begin{equation}
    \min_{\mathbf{b}, \mathbf{v}, \mathbf{w}} ~ J(\mathbf{b}, \mathbf{v}, \mathbf{w}; \lambda) ~ \mathrm{s.t.}~ w_{k} = 1.0, ~ \forall k: x_k \in S.
\end{equation}
We refer to this model as the participant-vocabulary consistency model because we optimize the consistency of scores computed based on the participants of each social interaction as well as the vocabulary used in each interaction.
The objective function \cref{eq:objective} is not jointly convex; However, if we optimize each parameter vector in isolation, we then solve convex optimizations with closed form solutions. The optimal value for each parameter vector given the others can be obtained by solving for their zero-gradient conditions. The update for bully score vector $\bf b$ is:
\begin{align*}
\argmin_{b_i} J = \frac{\displaystyle \sum_{m \in  M | s(m) = i }  \sum_{k \in f(m)} \left(w_k - |f(m)| v_{r(m)} \right)}{\displaystyle \lambda + \sum_{m \in  M | s(m) = i } |f(m)|}
\end{align*}
where $\{m \in M | s(m) = i\}$ is the set of messages that are sent by user $i$, and $|f(m)|$ is the number of n-grams in the message $m$. The closed-form solution for optimizing with respect to victim score vector $\bf v$ is
\begin{align*}
\argmin_{v_j} J = \frac{\displaystyle \sum_{ m \in M | r(m) = j } \sum_{k \in f(m)} \left( w_k - |f(m)| b_i \right)}{ \displaystyle \lambda + \sum_{m \in M | r(m) = j}|f(m)|} 
\end{align*}
where $\{ m \in M | r(m) = j \}$ is the set of messages sent to user $j$. The word score vector $\mathbf{w}$ can be updated with
\begin{align*}
\argmin_{w_k} J =  \frac{\displaystyle \sum_{m \in M | k \in f(m)}\left(b_{r(m)} + v_{s(m)}\right)}{\displaystyle \lambda + |\{m \in M | k \in f(m)\}|}.
\end{align*}
The set $\{m \in M | k \in f(m)\}$ indicates the set of messages that contain the $k$th feature or n-gram.
The strategy of fixing some parameters and solving the optimization problem for the rest of parameters known as alternating least-squares. Our algorithm iteratively updates each of parameter vectors $\mathbf{b}$, $\mathbf{v}$, and $\mathbf{w}$ until convergence.
The output of the algorithm is the bully and victim score of all the users and the bully score of all the words.

\section{Experiments}

\begin{figure}[tbp]
\vspace{-.7em}
\centering
\includegraphics[height = 45mm]{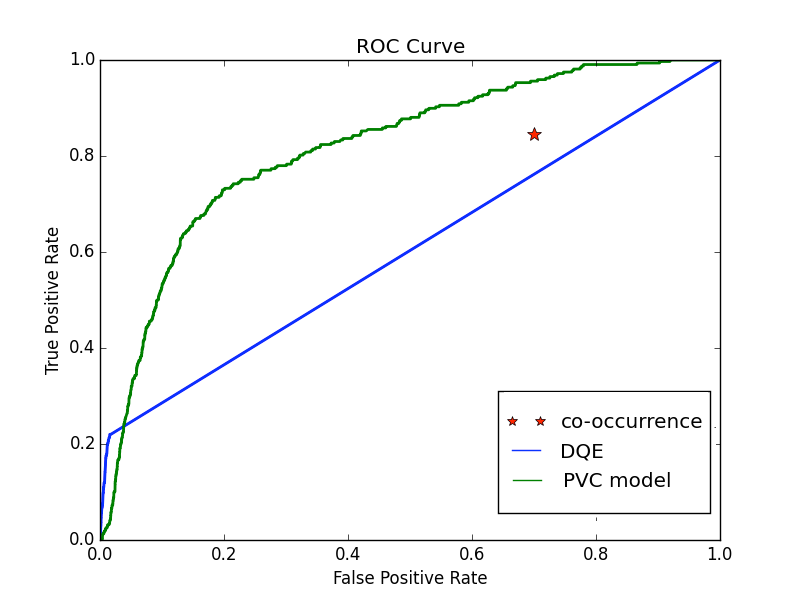}\\
\includegraphics[height = 45mm]{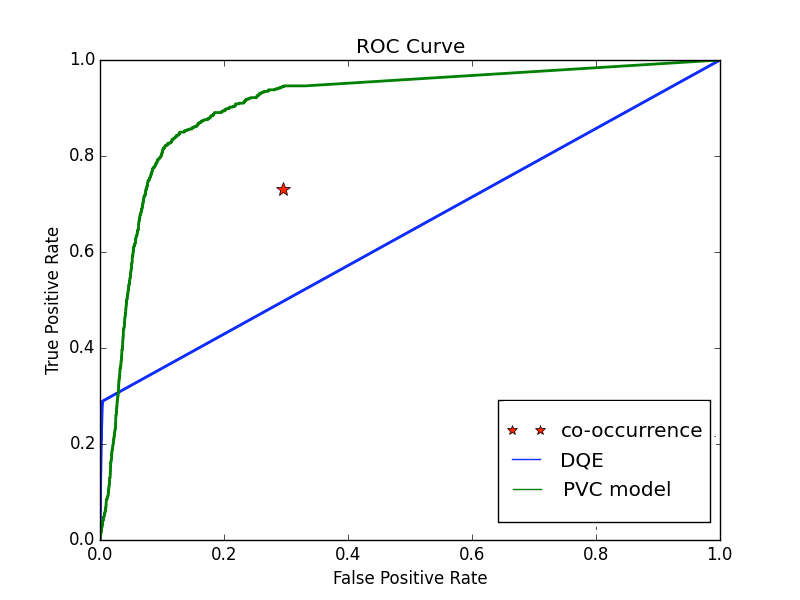}
\vspace{-1em}
\caption{ROC curve of recovered target words for Ask.fm (top) and Twitter (bottom).}
\vspace{-1.6em}
\label{roc}
\end{figure}

We ran our experiments on Twitter and Ask.fm data. To collect our Twitter data set, we first collected tweets using words from our curse word dictionary; then we use snowball sampling from these profane tweets to collect additional messages. Since we want to consider messages between users, we are interested in @-replies, in which one user directs a message to another or answers a post from another user. From our full data set, we remove the tweets which are not part of any such conversation, all retweets and duplicate tweets. After the data collection and post-processing, our Twitter data set contains 180,355 users and 296,308 tweets.

Ask.fm is a social question-answering service, where users post questions on the profiles of other users. We use part of the Ask.fm data set collected by \citet{hosseinmardi:BdCloud14}. They used snowball sampling, collecting user profiles and a complete list of answered questions. We remove all the question-answer pairs where the identity of the questioner is hidden. Our Ask.fm data set consists of 41,833 users and 286,767 question-answer pairs.


We compare our method with two baselines. The first is \textbf{co-occurrence}. All words or bigrams that occur in the same tweet as any seed word are given the score 1, and all other words have score 0. The second baseline is \textbf{dynamic query expansion} (DQE) \citep{ramakrishnan:kdd14}. DQE extracts messages containing seed words, then computes the \emph{document frequency score} of each word. It then iterates selection of the $k$ highest-scoring keywords and re-extraction of relevant messages until it reaches a stable state. 


\begin{table}[tbp]
    \begin{center}
\caption{Identified bullying bigrams detected by participant-vocabulary consistency from Twitter and Ask.fm data sets.}
    \begin{tabular}{ l  p{6cm} }
    \toprule
    \textbf{Data Set} & \textbf{Selected High-Scoring Words} \\ 
    \midrule
    Twitter &  sh*tstain~arisew, c*nt~lying, w*gger, commi~f*ggot, sp*nkbucket~lowlife, f*cking~nutter, blackowned~whitetrash, monster~hatchling, f*ggot~dumb*ss, *ssface~mcb*ober, ignorant~*sshat \\
     \addlinespace
	Ask.fm &  \mbox{total~d*ck}, blaky, \mbox{ilysm~n*gger}, fat~sl*t, pathetic~waste, loose~p*ssy, \mbox{c*cky~b*stard}, wifi~b*tch, \mbox{que*n~c*nt}, \mbox{stupid~hoee}, sleep~p*ssy, worthless~sh*t, \mbox{ilysm~n*gger}  \\ 
\bottomrule
    \end{tabular}
\label{tbl:t-swear}
\vspace{-1.5em}
\end{center}
    \end{table}

We evaluate performance by considering held out words from our full curse word dictionary as the relevant target words. We measure the true positive rate and the false positive rate, computing the receiver order characteristic (ROC) curve for each compared method.
\cref{roc} contains the ROC curves for both Twitter and Ask.fm. Co-occurrence only indicates whether words co-occur or not, so it forms a single point in the ROC space. Our PVC model and DQE compute real-valued scores for words and generate curves. DQE produces very high precision, but does not recover many of the target words. However, co-occurrence detects a high proportion of the target words, but at the cost of also recovering a large fraction of non-target words. PVC is able to recover a much higher proportion of the target words comparing DQE. PVC enables a good compromise between recall and precision.

We also compute the average score of target words, non-target words, and all of the words. If the algorithm succeeds, the average target-word score should be higher than the overall average. For both Twitter and Ask.fm, our proposed PVC model can capture target words much better than baselines. 
We measured how many standard deviations the average target-word score is above the overall average. For Twitter, PVC provides a lift of around 1.5 standard deviations over the overall average, while DQE only produces a lift of 0.242. We also observe the same behavior for Ask.fm: PVC learns scores that have a defined lift between the overall average word score and the average target word score (0.825). DQE produces a small lift (0.0099). Co-occurrence has no apparent lift.

By manually examining the 1,000 highest scoring words, we find many seemingly valid bullying words. These detected curse words include sexual, sexist, racist, and LGBT (lesbian, gay, bisexual, and transgender) slurs. \cref{tbl:t-swear} lists some of these high-scoring words from our experiments.

\begin{figure}[htbp]
\centering
\fbox{\includegraphics[height=2.9in]{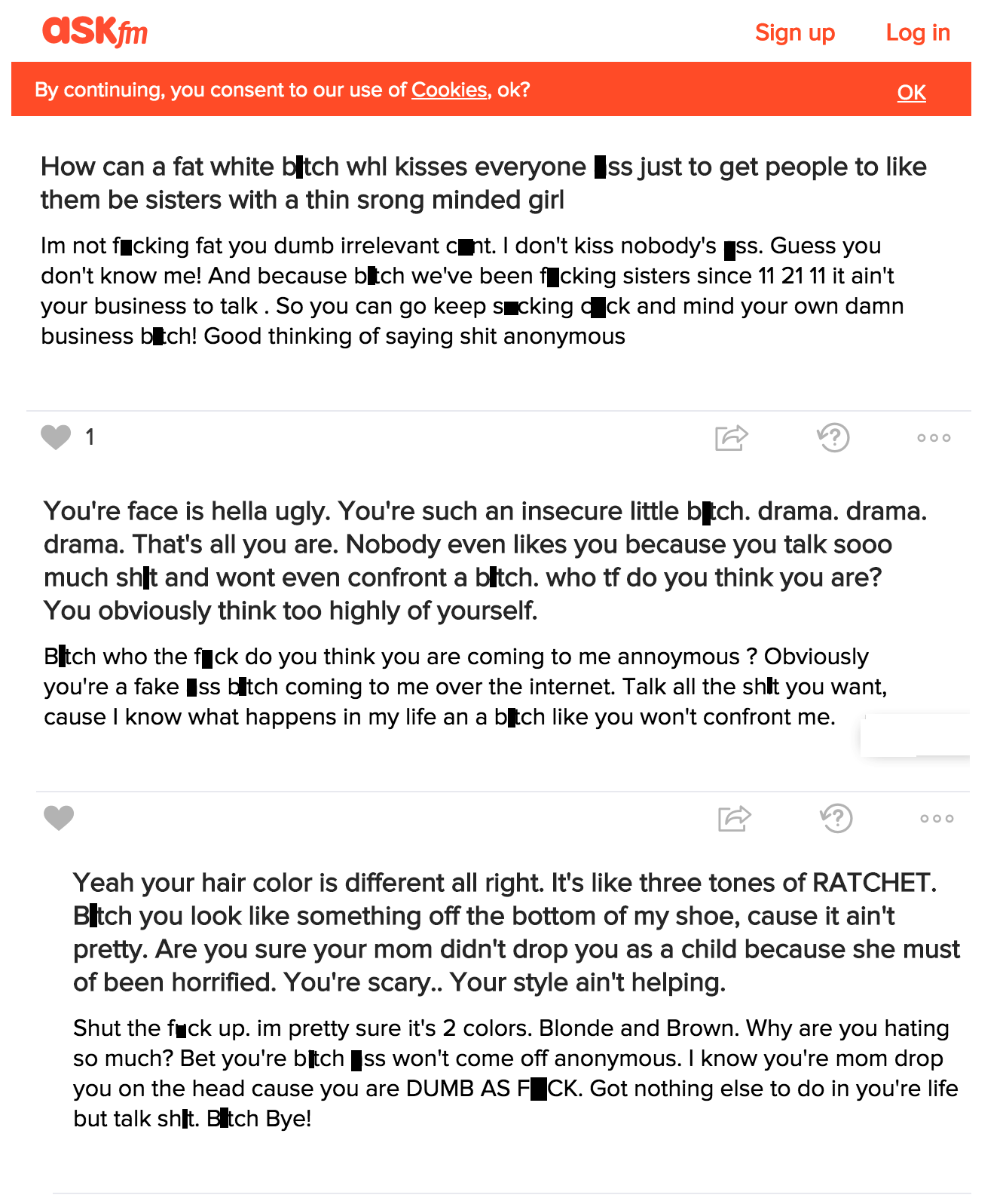}}
\vspace{-1em}
\caption{Example of an Ask.fm conversation containing possible bullying and heavy usage of offensive language.}
\vspace{-1em}
\label{fig:a1_ask}
\end{figure}

The PVC algorithm also computes bully and victim scores for users. By studying the profiles of highly scored victims in Ask.fm, we noticed that some of these users do appear to be bullied. This happens in Twitter as well, in which some detected high scoring users are often using offensive language in their tweets. \cref{fig:a1_ask} shows some bullying comments to an Ask.fm user and her responses, all of which contain offensive language and seem highly inflammatory.

\section{Conclusion}

In this paper, we proposed the participant-vocabulary consistency method to simultaneously discover victims, instigators, and vocabulary of words indicates bullying. Starting with seed dictionary of high-precision bullying indicators, we optimize an objective function that seeks consistency between the scores of the participants in each interaction and the scores of the language use. For evaluation, we perform our experiments on data from Twitter and Ask.fm, services known to contain high frequencies of bullying. Our experiments indicate that our method can successfully detect new bullying vocabulary. We are currently working on creating a more formal probabilistic model for bullying to robustly incorporate noise and uncertainty.

\bibliography{sigproc}
\bibliographystyle{icml2016}

\end{document}